\documentclass[twocolumn]{revtex4-1}
\usepackage{graphicx}                   
\usepackage[version=3]{mhchem}          
\usepackage[a4paper,plainpages=false,pdfstartview={XYZ null null 0.75}, colorlinks=true, citecolor=blue,linkcolor=blue]{hyperref}    
\usepackage[usenames,dvipsnames]{color}

\hypersetup{
    pdfauthor = {Roeland Juchtmans},
    pdftitle = {Properties of vortex beam diffraction},
    pdfsubject = {},
    pdfkeywords = {Vortex beam, electron vortex, diffraction}
}

\newcommand{\Int}[1]{\ensuremath{\displaystyle\int d{#1}\,}}

\newcommand{\Intbep}[3]{\ensuremath{\displaystyle \int_{#1}^{#2}}d{#3}\,}

\newcommand{\braket}[3]{\ensuremath{\left<{#1}\right|{#2}\left|{#3}\right>}}

\newcommand{\partieel}[2]{\ensuremath{\frac{\partial {#1}}{\partial {#2}}}}

\newcommand{\vt}[1]{\ensuremath{\boldsymbol{#1}}} 

\makeatletter

\newcommand{\Rmnum}[1]{\expandafter\@slowromancap\romannumeral #1@}
\makeatother

\begin{document}

\title{Extension of Friedel's law to Vortex Beam Diffraction}

\begin{abstract}
Friedel's law states that the modulus of the Fourier transform of real functions is centrosymmetric, while the phase is antisymmetric. As a consequence of this, elastic scattering of plane wave photons or electrons within the first-order Born-approximation as well as Fraunhofer diffraction on any aperture, is bound to result in centrosymmetric diffraction patterns. 
Friedel's law, however, does not apply for vortex beams, and centrosymmetry in general is not present in their diffraction patterns. 
In this work we extend Friedel's law for vortex beams by showing that
the diffraction patterns of vortex beams with opposite topological charge, scattered on the same two dimensional potential, always are centrosymmetric to one another, regardless of the symmetry of the scattering object. We verify our statement by means of numerical simulations and experimental data. Our research provides deeper understanding in vortex beam diffraction and can be used to design new experiments to measure the topological charge of vortex beams with diffraction gratings, or study general vortex beam diffraction.
\end{abstract}

\author{Roeland Juchtmans}
\affiliation{EMAT, University of Antwerp, Groenenborgerlaan 171, 2020 Antwerp, Belgium}
\author{Giulio Guzzinati}
\affiliation{EMAT, University of Antwerp, Groenenborgerlaan 171, 2020 Antwerp, Belgium}
\author{Jo Verbeeck}
\affiliation{EMAT, University of Antwerp, Groenenborgerlaan 171, 2020 Antwerp, Belgium}

\maketitle


\section{Introduction}

Friedel's law (FL) states that two antisymmetric points in the Fourier transform of a real function $V(\vt{r})$ are complex conjugated to one another, $\mathcal{F}[V](\vt{k})=\mathcal{F}^*[V](-\vt{k})$. By deriving this rule, the French crystallographer George Friedel was able to explain why, within the kinematical approximation, the zeroth order Laue zone in X-Ray diffraction patterns, always are centrosymmetric and showed how this puts restrictions on the possible crystal symmetries that can be directly determined by X-ray diffraction \cite{Friedel1913}.
Not only does FL apply to X-ray diffraction on crystals, it generally is valid for all scattering processes where a plane wave scatters on (approximately) two dimensional objects that are described by a real transmittance function, such as apertures or gratings.  Whereas for electron diffraction on crystals, FL breaks down relatively fast because of dynamical scattering, it remains valid for electrons scattered on apertures.
Since FL only is valid for plane waves, the diffraction patterns of distorted or modified beams, like the so-called vortex beams, do not necessarily follow this rule.

Vortex beams are eigenstates of the orbital angular momentum (OAM) operator, $\hat{L}_z=-i\hbar\partieel{}{\phi}$ \cite{Nye1974} and their wavefunction has the form
\begin{align}
\Psi_m(\vt{r})=\psi(r,z)e^{i\ell\phi},\label{eq:Vortex}
\end{align}
with $r$ and $\phi$ the radial and the azimuthal coordinate with respect to the wave propagation axis, $z$. The number $\ell$, is called the topological charge (TC) of the vortex. Since vortex beams are a eigenstates of the angular momentum operator, they possess a well-defined angular momentum of $\ell\hbar$ per photon\cite{Allen1999}, electron\cite{Bliokh2007} or any other particle\cite{Clark2015} described by the wave function in eq. \ref{eq:Vortex}. These beams have a typical donut-like intensity profile, a bright ring with a black spot in the middle, that arises from the phase singularity at its center.
Since their first experimental demonstration in optics \cite{Bazhenov1990}, they have been studied extensively, both theoretically and experimentally, which has led to numerous applications in fields such as nano-manipulation \cite{Luo,He,Friese}, astrophysics \cite{Foo2005,Swartzlander2007,Serabyn2010,Berkhout2008} and telecommunications \cite{andrewsbook,wangterabit,vortexnot}. 
Although slightly more complicated, a vast  amount of methods have been developed to produce vortex beams in an electron microscope \cite{Uchida2010,Verbeeck2010,Clark2013,Beche2013} as well and several studies suggest their use to probe magnetism \cite{Verbeeck2010,Juchtmans2016a}, nano-manipulation \cite{Verbeeck2013}, spin-polarization devices \cite{Karimi2012} and measuring the chirality of crystals \cite{Juchtmans2015}. Additionally, the expression in eq. \ref{eq:Vortex} provides an excellent mathematical basis to describe scattering on helical crystals \cite{Juchtmans2015a} or processes involving local exchange of OAM\cite{Juchtmans2015b}.
 
Many of the above applications, require an accurate measurement of the TC of the vortex beam. A popular way of doing this, is by studying the vortex diffraction patterns of specially designed apertures such as triangular apertures \cite{Hickmann2010,Guzzinati2014b} or multi-pinhole plates \cite{Guo2009,Shi2012,Clark2014}. In this work we elaborate on the observation made in these studies that diffraction patterns of vortex beams with opposite TC always seem to be centrosymmetric or equivalently, rotated $\pi$~rad, with respect to each other, independent from the shape of the aperture. Although in the above research, this can be explained using the symmetry of these specific apertures, we will formulate an extension of FL as a more fundamental property of vortex beam diffraction. We illustrate with numerical simulations and experiments, that even for non-symmetric scattering processes, the two-fold rotational symmetry between diffraction patterns with oppositely charged vortex beams, remains present. This research provides further insights in vortex beam diffraction and can help to optimize methods to measure the TC of vortex beams using diffraction gratings or study general vortex beam diffraction.

\section{Theoretical formulation}
\subsection{Friedel's law for plane waves}
Friedel's law (FL) includes two properties of the Fourier transform (FT) of real functions\cite{Friedel1913}.\\
Given a real function $f(\vt{r})$, its FT is given by
\begin{align}
\mathcal{F}(\vt{k})&=\Int{\vt{r}}{f(\vt{r})\,e^{i\vt{k}.\vt{r}}}.
\end{align}
Looking at the complex conjugate of the FT, 
\begin{align}
[\mathcal{F}(\vt{k})]^*&=\left[\Int{\vt{r}}{f(\vt{r})\,e^{i\vt{k}.\vt{r}}}\right]^*\nonumber\\&=\Int{\vt{r}}{f(\vt{r})\,e^{-i\vt{k}.\vt{r}}}=\mathcal{F}(-\vt{k}),\label{eq:CompConj}
\end{align}
one can easily see that the modulus of the FT is centrosymmetric,
\begin{align}
\left|F(\vt{k})\right|&=\left|F(-\vt{k})\right|\label{mod},
\end{align}
while its phase, $\phi(\vt{k})$, is antisymmetric,
\begin{align}
\phi(\vt{k})&=-\phi(\vt{k}),\label{arg}
\end{align}
Eq. \eqref{mod} and \eqref{arg} are known as Friedel's law and any set of centrosymmetric points $(\vt{k},-\vt{k})$ is called a Friedel pair.

Friedel applied these properties to describe the symmetries seen in X-ray diffraction patterns on crystals. For a scalar plane wave photon, that is scattered kinematically by a crysal, the scattering potential is a real function. Within the single scattering approximation, valid for weakly interacting samples, the scattering amplitude of a plane wave photon, with wavenumber $\vt{k}$, to scatter to a plane wave with wavenumber $\vt{k}'$, then is given by 
\begin{align}
A(\vt{k},\vt{k}')&=\braket{\vt{k}'}{V(\vt{r})}{\vt{k}}=\Int{\vt{r}}e^{-i\vt{k}'.\vt{r}}V(\vt{r})\,e^{i\vt{k}.\vt{r}}\nonumber\\
&=\Int{\vt{r}}{V(\vt{r})\,e^{i\Delta \vt{k}.\vt{r}}}\label{eq:ScattAmp}
\end{align}
with $\Delta\vt{k}=\vt{k}-\vt{k}'$. Clearly, the scattering amplitude exactly equals the FT of the potential. The X-ray diffraction pattern is given by the intersection of the three dimensional scattering amplitude with the so-called Ewald sphere, which assures that the length of the wave vector of the outgoing photon, and thus the energy, is the same as that of the incoming photon \cite{DeGraef2003}. In general, the curvature of this sphere can be considered flat for any relevant scattering angle, and the scattering amplitude is determined by the two dimensional FT of the potential projected along the $z$-axis, the photon's propagation axis, $V_{\perp}(\vt{r}_{\perp})=\Int{z}{V(x,y,z)}$,
\begin{align}
A(\vt{k},\vt{k}')=\Int{\vt{r}_{\perp}}{V_{\perp}(\vt{r}_{\perp})\,e^{i \vt{k}'_{\perp}.\vt{r}_{\perp}}},
\end{align}
where here, and in the following, $\vt{r}_\perp$ denotes the two dimensional coordinate in the $(x,y)$ plane. The diffraction patterns in the kinematical approximation, given by the scattering amplitude squared $|A(\vt{k},\vt{k}')|^2$, will be centrosymmetric because of FL, making it impossible to distinguish certain symmetries in the projection of crystals with one single diffraction pattern. A crystal with a three-fold rotation axis along the projection direction, for instance, will show the same symmetry in its diffraction pattern as a crystal with a six-fold rotation axis \cite{Friedel1913}. 

Besides X-ray diffraction on crystals, FL also applies to diffraction of plane waves on apertures and gratings.
Consider, for instance, an aperture in the $xy$-plane that is placed at $z_0=0$, illuminated by a plane wave of the form $\Psi_0= e^{ik_z.z_0}$. The diffracted wave in the far--field of the aperture then is given by the Fraunhofer equation \cite{Born1999}
\begin{align}
\Psi(\vt{k}_\perp,z)&=\Int{\vt{r}'_\perp}{f(\vt{r}'_\perp)e^{i\frac{k_z}{z}\vt{r}_\perp \cdot \vt{r}'_\perp}}\Psi_(\vt{r}'_\perp,0)\nonumber\\
&=\Int{\vt{k}'_\perp}{f(\vt{r}'_\perp)e^{i\frac{k_z}{z}\vt{r}_\perp \cdot \vt{r}'_\perp}}\nonumber\\
&=\Int{\vt{k}'_\perp}{f(\vt{r}'_\perp)e^{i\vt{k}_\perp \cdot \vt{r}'_\perp}},\label{eq:Fraunhofer}
\end{align}
with $\vt{k}_\perp=\frac{k_z}{z}\vt{r}'_\perp$, $z$ te distance from the aperture and $f(x',y')$ the transmittance function of the grating. In general $f(\vt{r}'_\perp)$ does not have to be real. In case of phase plates, for instance, where the phase of the wave is altered, $f(\vt{r}'_\perp)$ is complex. But for apertures that only change the amplitude, $f(\vt{r}'_\perp)$ is real. In this case it is clear we can apply Friedel's law and the intensity pattern of the diffracted wave will always be centrosymmetric, independently from the symmetry of the diffraction grating. This is shown numerically and experamentally in fig. \ref{fig:intcenter}.

\subsection{Friedel's law for vortex beams}

Friedel's law only applies to diffraction of plane waves. When the incoming beam is modified or distorted, it is no longer valid. This is the case for an incoming vortex beam of the form
\begin{align}
\Psi_\ell(\vt{r})=\psi_\ell(r,z)e^{i\ell\phi},
\end{align}
where $\ell$ is the topological charge and $\psi_\ell(r,z)$ is the radial profile of the beam at height $z$. The symmetry of the diffraction pattern of such a beam scattered on an aperture or crystal in general will not be centrosymmetric anymore, see fig. \ref{fig:intcenter}, fig. \ref{fig:tricenter} and fig. \ref{fig:trioffcenter}. However, applying the same trick as in eq. \ref{eq:CompConj}, we can easily find a relation between the diffraction patterns of vortex beams with opposite topological charge having the same \emph{real} radial profile. 

Consider scattering of a vortex beam on an aperture with real transmittance function $f(\vt{r}_\perp)$ placed at height $z_0=0$. Similar to eq. \ref{eq:Fraunhofer}, the wave in the far--field now is given by the Fraunhofer equation 
\begin{align}
\Psi_\ell(\vt{r}_\perp,z)&=\Int{\vt{r}'_\perp}{f(\vt{r}'_\perp)e^{i\frac{k_z}{z}\vt{r}_\perp \cdot \vt{r}'_\perp}}\Psi_\ell(\vt{r}'_\perp,0).
\end{align}
When we look at the complex conjugate of the scattered wave, we get
\begin{align}
\Psi_\ell^*(\vt{r}_\perp,z)&=\left(\Int{\vt{r}'_\perp}{f(\vt{r}'_\perp)e^{i\frac{k_z}{z}\vt{r}_\perp \cdot \vt{r}'_\perp}}\Psi_\ell(\vt{r}'_\perp,0)\right)^*\nonumber\\
&=\Int{\vt{r}'_\perp}{f(\vt{r}'_\perp)e^{-i\frac{k_z}{z}\vt{r}_\perp \cdot \vt{r}'_\perp}}\psi_\ell(r,0)e^{-i\ell\phi}\nonumber\\
&=\Psi_{-\ell}(-\vt{r}_\perp,z)\label{eq:FriedelVortex}
\end{align}

This means that diffraction patterns of oppositely charged vortex probes with a real radial profile, will be centrosymmetric, or rotated $\pi$~rad, with respect to each other. The same goes for diffraction on a crystal when replacing the incoming plane wave in eq. \ref{eq:ScattAmp} by a vortex beam. Note that the radial profile of the beam has to be real, as is the case for most typical vortex beams such as the non-diffracting Bessel beams, $\psi^{\text{B}}_\ell(r,z)=J_\ell(kr)$, or the Laguerre Gaussian beams in the focal plane, $\psi^{\text{LG}}_\ell(r,0)=\left(\frac{r}{k}\right)^\ell L^\ell_n\left(\frac{r^2}{k^2}\right)\exp\left(-\frac{r^2}{k^2}\right)$, where $L_n^\ell$ is the Laguerre polynomial with radial and angular mode $n$ and $\ell$ respectively and $k$ is a parameter determining the with of the beam. Note that when defocussing vortex beams with a real radial function, it becomes complex and eq. \ref{eq:FriedelVortex} no longer holds.

We will refer to eq. \ref{eq:FriedelVortex} as the extension of Friedel's law (EFL) applicable for vortex beams. It includes the known Friedel's law when considering a plane wave as a vortex beam with topological charge $\ell=0$.

\section{Simulation and experiment}
In this section, we show simulations of vortex beam diffraction on a multi-pinhole and a triangular aperture and compare these with electron vortex beam experiments on a probe-corrected FEI Titan$^3$ microscope. The incoming vortex wave with TC, $\ell$, numerically is simulated by:
\begin{align}
\Psi_\ell(r,\phi)&=\Intbep{0}{k_{max}}{k}\Intbep{0}{2\pi}{\phi_k}e^{i\ell\phi_k}e^{ikr\cos(\phi_k-\phi)},\label{vortexbeam}
\end{align}
with $k_{max}=\alpha_ck_0$, $k_0=\frac{2\pi}{\lambda}$. Here $\lambda$ is the wavelength of the incoming beam and $\alpha_c$ is the convergence angle, an experimental parameter determining the spot size. In our simulations and experiments we make use of a 300~keV electron beam, $k_0=3.19$~pm$^{-1}$ with a semi--convergence angle $\alpha_c=4$~$\mu$rad. Eq. \ref{vortexbeam} reflects the experimental setup of a vortex beam that is generated by evenly illuminating the condenser aperture while giving it an extra vortex phase, after which the probe is formed in the far--field of this aperture. For photons, this phase can be applied with, for instance, a spatial light modulator \cite{Bouchal2005}, while for electrons, this can be realized by placing the tip of a magnetized needle on the center of the condenser aperture \cite{Beche2013}. In our experiment, however, the electron vortex beams are created by placing a fork hologram in the illumination system \cite{Verbeeck2010}. The condenser system then is used to project the FT of the hologram in the sample plane and the vortex probe in eq. \ref{vortexbeam} is obtained. For practical reasons, the triangular and multi-pinhole apertures were placed in the selected area plane which is conjugated to the image plane. Note that the experiments are done using an electron microscope, but the result is equally valid for optical vortex diffraction.

\begin{figure}[t!]\begin{center}
		\includegraphics[width=\columnwidth]{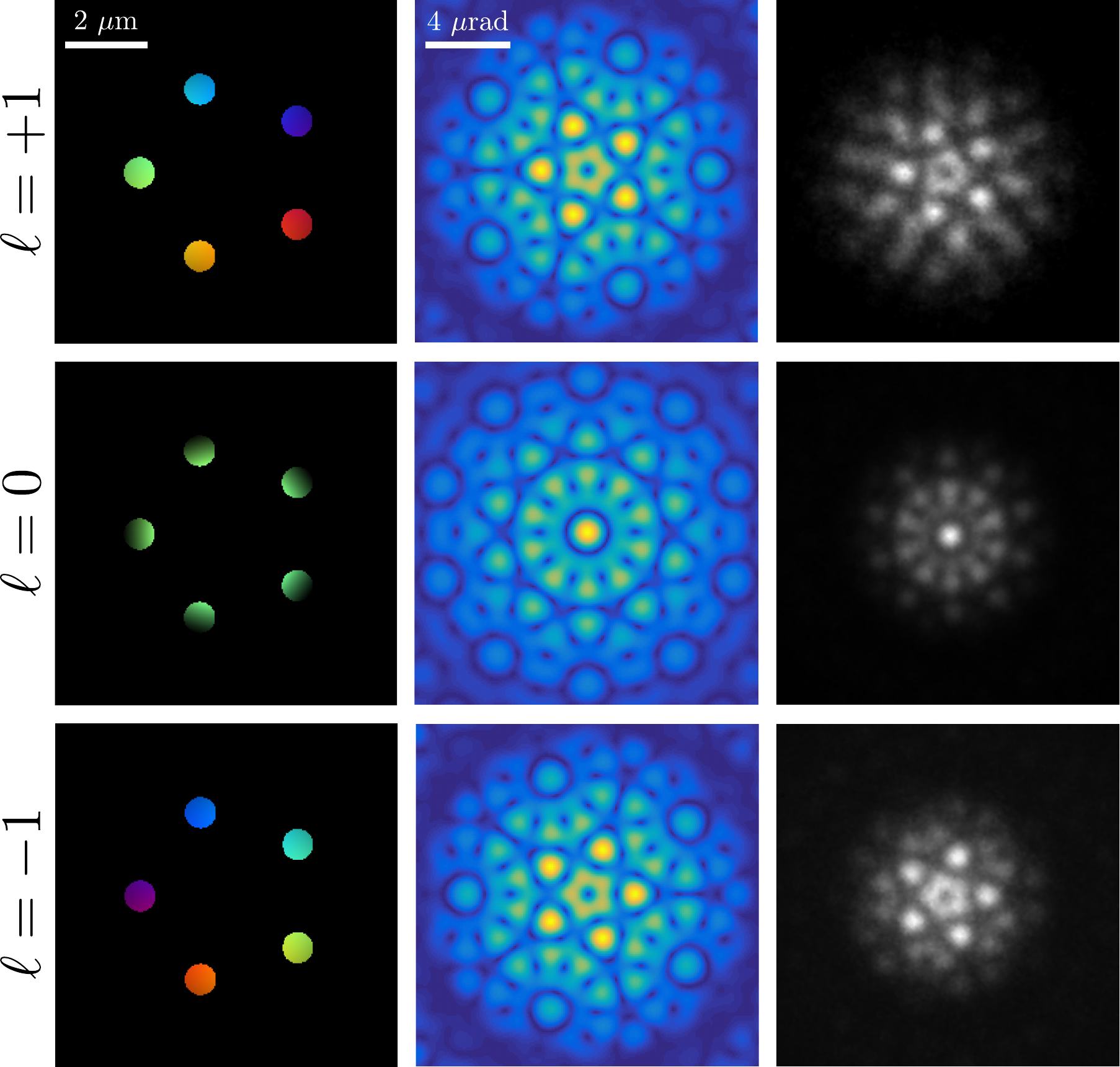}
	\end{center}
	\caption{(Left) Illumination of a five-fold multi-pinhole aperture with an $\ell=-1$, $\ell=0$ and $\ell=1$ vortex, as in \cite{Guo2009} and \cite{Clark2014}. (Center) Simulated diffraction pattern. (Right) Experimental diffraction pattern. Since the axis of the vortex lies on the five-fold symmetry-axis of the aperture, all diffraction patterns show the five-fold symmetry of the pinhole aperture. The $\ell=0$ diffraction pattern is two-fold rotation symmetric as well, demonstrating the conventional Friedel's law. This symmetry is absent in the $\ell=-1$ and $\ell=1$ diffraction patterns, but, as expected from eq. \ref{eq:FriedelVortex}, they do show two-fold rotational symmetry with respect to each other.\label{fig:intcenter}}
\end{figure}

\begin{figure}[thbp]\begin{center}
\includegraphics[width=\columnwidth]{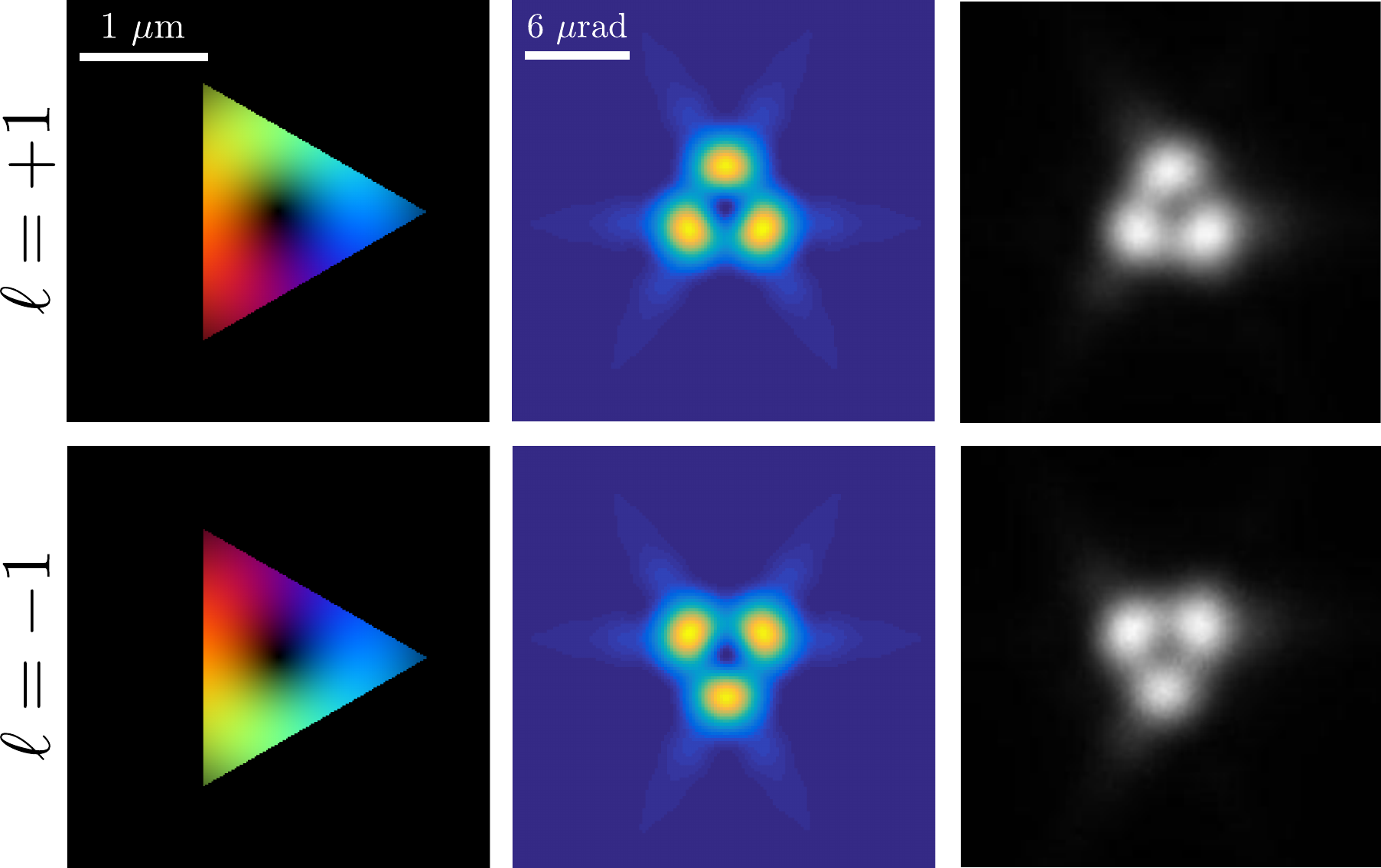}
        \end{center}
\caption{(Top) Illumination of a triangular aperture with an $\ell=-1$ and $\ell=1$ vortex, as in \cite{Hickmann2010}. (Middle) Simulated diffraction pattern. (Bottom) Experimental diffraction pattern. Since the axis of the vortex lies on the three-fold symmetry-axis of the aperture, all diffraction patterns show three-fold symmetry. Again, the $\ell=-1$ and $\ell=1$ diffraction patterns, show two-fold rotational symmetry with respect to each other.\label{fig:tricenter}}
\end{figure}

In fig. \ref{fig:intcenter}, the real space image of a vortex beam with $\ell=-1$, $\ell=0$ and $\ell=+1$ centered on a multi-pinhole aperture that consists out of five equidistant holes, is shown together with the simulated and experimental diffraction pattern. Because of the five-fold symmetry of the aperture and the central position of the probe, the diffraction patterns all show five-fold symmetry. Also, because of Friedel's law, the $\ell=0$ beam shows extra two-fold rotation symmetry and the resulting diffraction pattern becomes ten-fold symmetric. This symmetry is absent in the vortex beam diffraction patterns. However, following eq. \ref{eq:FriedelVortex}, a two-fold rotation symmetry can be observed between the $\ell=-1$ and $\ell=+1$ diffraction patterns.
The same can be seen in fig. \ref{fig:tricenter}, where a $\ell=-1$ and an $\ell=+1$ vortex are centered on a triangular aperture. Because of the three-fold symmetry of the aperture and the central probe position, both diffraction patterns show three-fold symmetry. Again, the two diffraction patterns are rotated $\pi$~rad with respect to each other. 

\begin{figure}[tp]\begin{center}
		\includegraphics[width=\columnwidth]{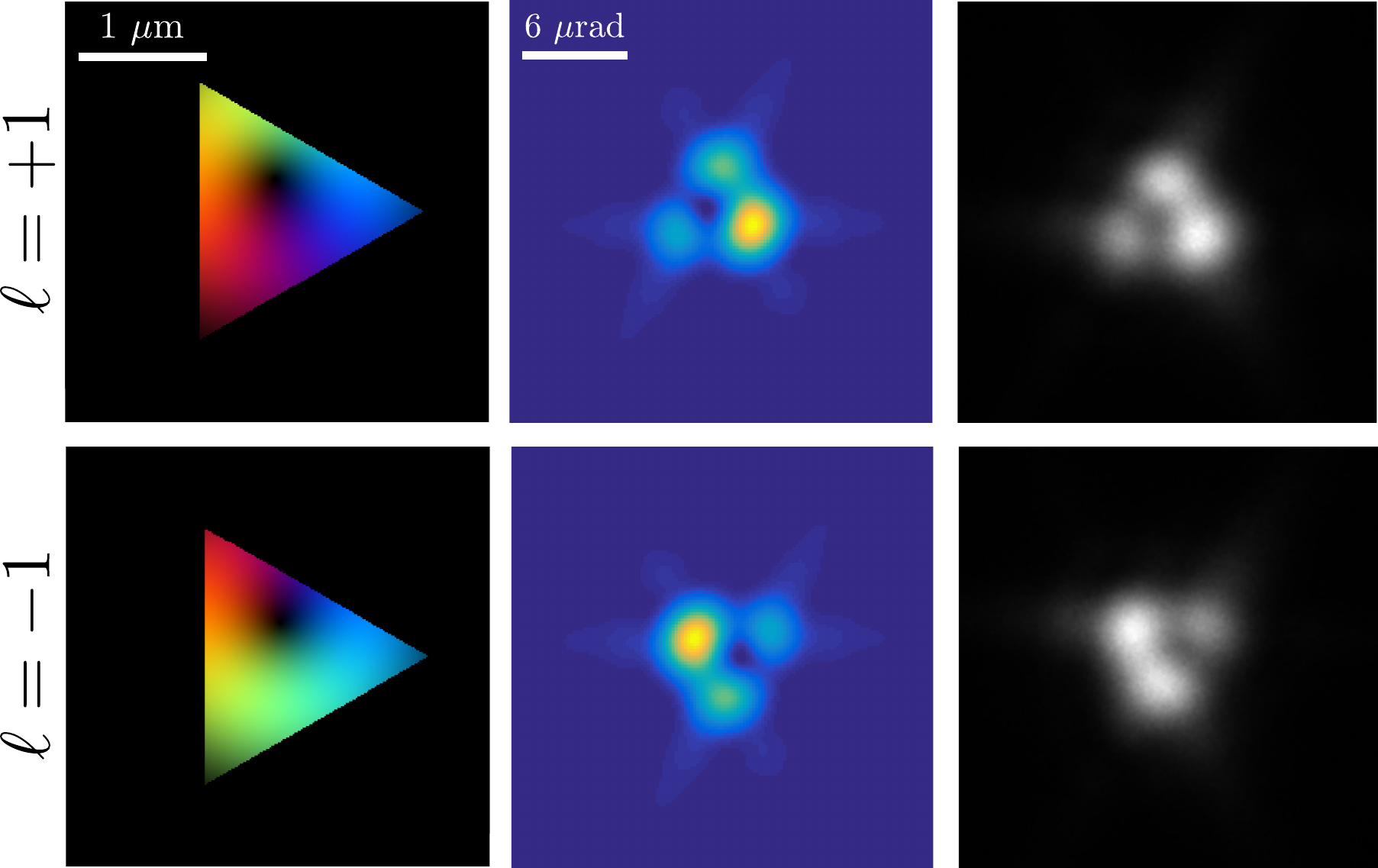}
	\end{center}
	\caption{Same as in fig. (\ref{fig:tricenter}) but now the axis of the vortex lies on a non symmetric point of the aperture. Therefore the diffraction patterns no longer show the symmetry of the aperture. However, consistent with eq. \ref{eq:FriedelVortex}, the $\ell=-1$ and $\ell=1$ diffraction patterns are rotated $\pi$~rad with respect to one another. This shows that this effect is not dependent on the symmetry of the scattering object, but is a more fundamental effect of vortex scattering.\label{fig:trioffcenter}}
\end{figure}

In both examples, the vortex is centered on a high-symmetry point of the aperture, which makes that the symmetry of the aperture is retained in all diffraction patterns. This makes that the two-fold rotational relation between the opposite vortex diffraction patterns, is equivalent with a mirror operation or a rotation over $\pi/5$ and $\pi/3$~rad for the five-fold and three-fold symmetric aperture respectively. To show that the two-fold rotation is independent from the symmetry of the beam position and symmetry of the aperture, in fig. \ref{fig:trioffcenter}, we shift the vortex beam to a non symmetric position on the triangular aperture, thereby destroying the three-fold symmetry. However, the two-fold rotation relation between diffraction patterns of the opposite vortex beams, remains present as expected.

\section{Discussion}

The EFL in eq. \ref{eq:FriedelVortex} shows that diffraction patterns of oppositely charged vortices always are two-fold symmetric with respect to each other when scattered on a real two-dimensional projected potential or two-dimensional apertures with real transmittance functions. The triangular aperture we used, fig. (\ref{fig:tricenter}) already was suggested to determine the topological charge of a vortex beam by Hickmann \emph{et al.}~\cite{Hickmann2010}. By looking at the diffraction pattern, they were able to derive a rule for determining the magnitude of OAM of the incoming beam and they noted that diffraction patterns of opposite charge always were rotated by $\pi$~rad. Similarly, the pinhole aperture was proposed by Berkhout \emph{et al.} \cite{Berkhout2008} as an alternative method to measure the topological charge of vortex beams. They also mention that, in the case of an odd number of holes, the diffraction pattern of opposite charge are mirrored to one another. 
Several other diffraction experiments with vortices were performed \cite{Sztul2006,Guo2009,Mazilu2012,Ghai2009,Shi2012,Liu2013a,Ferreira2011} and, when a comparison with oppositely charged vortices was made, the same centrosymmetry between diffraction patterns always was observed.

In all these examples, however, the arguments to explain the symmetry between two opposite vortex beam diffraction patterns, relied on the symmetry of the specific apertures with respect to the center of the vortex (the studies on asymmetric apertures we found, unfortunately, do not compare opposite vortex beam diffraction patterns). The EFL, however, shows that this symmetry is a fundamental property of vortex beams scattered on any real, projected potential or two-dimensional apertures with real transmittance functions, independent from the symmetry of the scattering object.

Like Friedel's law, EFL is useful when studying crystallography with vortex beams. In previous work, for example, we proposed the use of electron vortex beam diffraction to determine the handedness of chiral crystals \cite{Juchtmans2015}, crystals that only differ by a mirror operation,  starting from the idea that vortex beams themselves are chiral. However, by extending Friedel's law to vortex beams, it immediately becomes clear that the chirality can-not be learned directly from the projected potential of the crystal or the zeroth order Laue zone (ZOLZ) in the diffraction pattern, within the kinematical approximation. Any chiral effect that would be seen with one vortex, would only differ by a rotation of $\pi$~rad, with the opposite vortex. Therefor the left-handed crystal would show the same ZOLZ as its right-handed mirror image when rotated $\pi$~rad along the beam's propagation direction. Looking at kinematically scattered electrons, any chiral effect must thus be searched for in the higher order Laue zones that contain three dimensional information about the crystal\footnote{In this case, it is easy to show that the HOLZ of a $\ell$ vortex beam is rotated $\pi$~rad with respect to the $-\ell$ vortex, when, at the same time, the potential is mirrored along the plane perpendicular to the propagation of the vortex beam.}.

\section{Conclusion}

In this work, we derived an extension of Friedel's law (EFL) that applies to vortex beams, whether it concerns an optical, electron or any other particle beam. We show that when a vortex beam is diffracted on a two-dimensional scattering object, such as apertures, two diffraction patterns of vortex beams with opposite topological charge, always are rotated $\pi$~rad with respect to  each other. This is independent from the symmetry of the scattering object and a fundamental property of vortex beam scattering.  Our findings also apply for diffraction on a crystal, when kinematical scattering between the zeroth order Laue zone is considered. In general, this can be applied to X-ray, but not for electron diffraction on crystals, which is mostly dominated by dynamical scattering.
  
We verified our analytical derivation with numerical simulations and showed experimental results for electron vortex beams scattered on a multi-pinhole and triangular aperture. We compared our findings with observations made in literature, where this effect mostly is explained using the symmetry of the scattering objects with respect to the center of the vortex beam. However, with our derivation and by looking at vortex beams scattering on non-symmetric points of our apertures, we've shown that this symmetry is a more fundamental property of vortex beam diffraction. The work presented here provides deeper understanding in vortex beam diffraction, which in turn can be used to design new experiments to measure the topological charge of vortex beams with diffraction gratings, or study general vortex beam diffraction.

\section{acknowledgments}
The authors acknowledge support from the FWO (Aspirant Fonds Wetenschappelijk Onderzoek - Vlaanderen) and the EU under the Seventh Framework Program (FP7) under a contract for an Integrated Infrastructure Initiative, Reference No. 312483-ESTEEM2 and ERC Starting Grant 278510 VORTEX.

\end{document}